\documentclass[reprint,superscriptaddress,citeautoscript,floatfix,nofootinbib]{revtex4-1}%
\usepackage{latexsym}
\usepackage{graphics}
\usepackage{amsmath,amssymb,amsfonts,bbm,textcomp}
\usepackage[english]{babel}
\usepackage[utf8]{inputenc}
\usepackage[T1]{fontenc}
\usepackage{times}
\usepackage[shortcuts]{extdash}
\usepackage{hyperref}
\hypersetup{pdftitle=Fractional quantization of charge and spin in topological quantum pumps, pdfauthor=Pasquale Marra and Roberta Citro}

\newcommand{\up}{{\uparrow}}
\newcommand{\down}{{\downarrow}}
\newcommand{\nodag}{{\phantom{\dag}}}

\begin{document}

\title{Fractional quantization of charge and spin in topological quantum pumps}
%\author{Pasquale Marra\inst{1,2}\mail{pasquale.marra@spin.cnr.it} \and Roberta Citro\inst{2,1}}%
%\institute{CNR-SPIN, I-84084 Fisciano (Salerno), Italy \and Department of Physics ``E.\,R.\, Caianiello'', University of Salerno, I-84084 Fisciano (Salerno), Italy}%
\author{Pasquale Marra}\email{pasquale.marra@spin.cnr.it}\affiliation{CNR-SPIN, I-84084 Fisciano (Salerno), Italy}\affiliation{Department of Physics ``E.\,R.\, Caianiello'', University of Salerno, I-84084 Fisciano (Salerno), Italy}%
\author{Roberta Citro}\affiliation{Department of Physics ``E.\,R.\, Caianiello'', University of Salerno, I-84084 Fisciano (Salerno), Italy}\affiliation{CNR-SPIN, I-84084 Fisciano (Salerno), Italy}%

%\date{\today}

\begin{abstract}%
%\abstract{%
Topological quantum pumps are topologically equivalent to the quantum Hall state:
In these systems, the charge pumped during each pumping cycle is quantized and coincides with the Chern invariant.
However, differently from quantum Hall insulators, quantum pumps can exhibit novel phenomena such as the fractional quantization of the charge transport, as a consequence of their distinctive symmetries in parameter space.
Here, we report the analogous fractional quantization of the \emph{spin} transport in a topological spin pump realized in a one-dimensional lattice via a periodically modulated Zeeman field.
In the proposed model, which is a spinfull generalization of the Harper-Hofstadter model, the amount of spin current pumped during well-defined fractions of the pumping cycle is quantized as fractions of the spin Chern number.
This fractional quantization of spin is topological, and is a direct consequence of the additional symmetries ensuing from the commensuration of the periodic field with the underlying lattice.
%
%\PACS{%
%{73.43.-f}{Quantum Hall effects}	\and %
%{72.25.-b}{Spin polarized transport}	\and %
%{37.10.Jk}{Atoms in optical lattices} %
%%03.65.Vf Phases: geometric; dynamic or topological
%%73.21.Cd Superlattices
%%67.85.-d Ultracold gases, trapped gases
%     }%
%} %
\end{abstract}%

%\titlerunning{Fractional quantization of charge and spin in topological quantum pumps}%
%\authorrunning{Pasquale Marra and Roberta Citro}%
\maketitle

\section{Introduction}

Topological insulators are characterized by the presence of gapless edge modes which are topologically protected by a nontrivial topological invariant\cite{Hasan2010,Qi2011}.
In particular, quantum Hall insulators\cite{Klitzing1980} exhibit chiral edge modes and a quantized Hall conductance, which coincides with the Chern invariant, i.e., an integer capturing the global topological properties of the system\cite{Thouless1982}.
On the other hand, the more recently discovered quantum spin Hall insulators\cite{Hasan2010,Qi2011} exhibit spin-polarized helical edge modes which are protected by a time-reversal symmetric $\mathbb{Z}_2$ topological invariant or by a spin Chern number\cite{Kane2005Graphene,Kane2005,Bernevig2006,Sheng2006}.
The quantum \emph{spin} Hall state is equivalent to two copies of the \emph{charge} Hall state cloned into two decoupled spin channels, with spin up and down electrons moving in opposite directions.
As such, the total charge Hall conductance vanishes but the net spin Hall conductance remains finite\cite{Bernevig2006}.

As shown by Thouless\cite{Thouless1983,Niu1984}, a one-dimensional system with a periodically-modulated field can realize a topological quantum pump. 
This system can exhibit a topologically nontrivial state which is equivalent to a two-dimensional quantum Hall insulator.
In this state, the charge pumped during each pumping cycle is quantized and coincides with the Chern invariant:
The quantized charge is topological, being analogous to the transverse conductance of the quantum Hall state.
As a further generalization, the topological $\mathbb{Z}_2$ spin pumping and spin Chern pumping has been theoretically proposed\cite{Fu2006}. 
These are indeed the one-dimensional analogous of a spin quantum Hall state, in view of the fact that the spin pumped is finite despite a vanishing net charge pumped during each cycle.
Yet, topological spin pumps remain difficult to realize in condensed matter settings, despite the numerous proposals\cite{Shindou2005,Citro2011,Mei2012,Ferraro2013,Zhou2014,Deng2015,MNChen2015}.
On the other hand, ultracold atomic gases in optical lattices have recently proven as an ideal platform to realize exotic states of matter\cite{Lewenstein2007,Bloch2008,Dalibard2011}.
In particular, recent advances in generating artificial magnetic fields and spin-orbit coupling in neutral atoms\cite{Aidelsburger2011,JimenezGarcia2012,Struck2012,Kartashov2013,Goldman2014} have brought to the experimental realization of the Hofstadter model\cite{Lang2012,Aidelsburger2013,Miyake2013,Aidelsburger2015}
and of topological charge\cite{Lohse2016,Nakajima2016,Citro2016} and spin\cite{Schweizer2016} pumps in optical lattices.
On top of that, topological quantum pumps and, in general, low-dimensional systems with periodically-modulated fields, can exhibit novel and characteristic physical phenomena which have no analogue in quantum Hall insulators\cite{Lopes2016,Zeng2016,Ronetti2017}, such as the presence of fractional edge charges\cite{Gangadharaiah2012,Park2016}, and the fractional quantization of the charge transport\cite{Marra2015}, which is a direct consequence of their peculiar symmetries in parameter space.
Indeed, in the presence of a periodically modulated field, the charge pumped at well-defined fractions of the pumping period is quantized as integer fractions of the Chern number, as shown in Ref.~\onlinecite{Marra2015}.

In this work we describe a novel but related physical phenomenon, i.e., the fractional quantization of the spin transport in a spin quantum pump.
This system can be realized by a one-dimensional lattice in the presence of a spatially and periodically modulated Zeeman field, with a periodicity which is commensurate with the underlying lattice, and which is modulated adiabatically in time.
This non-interacting quantum spin pump exhibits a fractional quantization of the amount of spin pumped at well-defined fractions of the pumping cycle.
In particular, we show that the fractional quantization of the spin transport can be probed by measuring the variations of the spin center of masses of the electronic gas.
We therefore discuss the experimental implementation of this system in condensed matter systems as well as in ultracold atomic gases in optical lattices, which have been proven as an ideal platform to realize exotic states of matter\cite{Lewenstein2007,Bloch2008,Dalibard2011,Zhang2012,Zhang2012PRA,Mei2012,Lang2012,Zhang2015,Zhang2016}.

\section{Fractional quantization of charge and spin transport}
\label{sec:general}

Quantum pumps are defined as insulating systems which can transfer a finite amount of charge or spin as a result of the adiabatic and periodic evolution in time of a driving field.
Let us consider the case where the adiabatic evolution of the system is described by a slow and continuous change of a parameter $\varphi$, and assume that the system is periodic in such parameter with period $2\pi$.
If the Fermi level lays in any of the intraband gaps of the spectrum, i.e., any energy gap which remains open for any choice of the control parameter $\varphi$, and the up and down spin channels are completely decoupled, the charge\cite{Thouless1983} $Q$ and the total spin\cite{Kane2005,Bernevig2006} $S$ pumped during an adiabatic cycle $\varphi_0\to\varphi_0+2\pi$ corresponds with the charge and the spin Chern numbers and are given by
\begin{align}
Q&=C_\up+C_\down,
\nonumber\\
S&=C_\up-C_\down,
\nonumber\\
\text{where \qquad}
C_\tau&=
\frac{1}{2\pi}
\sum_{i}
\int_{\varphi_0}^{\varphi_0+2\pi}
\hspace{-2em}
\mathrm{d}\varphi \int_\text{BZ}
\hspace{-.6em}
\mathrm{d}k \hspace{.2em}
\Omega_{i\tau}(k,\varphi).
\label{eq:ChargeSpinPump}
\end{align}
Here $C_\tau$ are the Chern numbers separately for the up and down spin channels $\tau=\up,\down$.
They are defined as the sum over all the filled bands of the integrals of the Berry curvature\cite{Hasan2010,Qi2011,Xiao2010} $\Omega_{i\tau}(k,\varphi)=\partial_k A_{i\tau}(\varphi) - \partial_\varphi A_{i\tau}(k)$, where $A_{i\tau}(\kappa)=\imath \langle i\tau|\partial_\kappa|i\tau\rangle$ is the Berry connection of the eigenstate $|i\tau\rangle$ of the $i$th band with spin $\tau$.
Therefore the charge $Q$ and the spin $S$ pumped during an adiabatic cycle are quantized, being respectively equal the sum and the difference of the Chern numbers of each spin channels, and do not depend on the initial phase $\varphi_0$.
Notice that the parameter $\varphi$ plays the role of an additional synthetic (non-spatial) dimension\cite{Kraus2013,Celi2014,Marra2015,Marra2016b}.
The one-dimensional physical system is thus embedded in a two-dimensional parameter space, which is thus analogous to an insulating two-dimensional system.
Topological charge pumps (i.e., $Q\neq0$) are therefore analogous to a quantum Hall insulator, whereas topological spin pumps (i.e., $S\neq0$) are analogous to a quantum spin Hall insulator.

For arbitrary adiabatic transformations with $\varphi_0\to\varphi_0+\Delta\varphi$ with $\Delta\varphi\neq2\pi$ instead, the charge and spin pumped are not quantized and depend in general on the initial phase $\varphi_0$.
Nevertheless, if the system exhibit an additional unitary periodicity in the control parameter $\varphi$, the charge and the spin pumped over fractions of the pumping cycle are quantized as fractions of the charge and spin Chern number.
To be more specific, consider the case where the Hamiltonian of the system is periodic up to a unitary transformation $U$ with respect to the control parameter $\varphi$ with period $2\pi/q$, that is
\begin{equation}
H(k,\varphi+2\pi/q)=U H(k,\varphi) U^\dagger,
\label{eq:Unitary}
\end{equation}
with $q$ integer.
In this case the Berry curvature, which is gauge invariant, is a periodic function of the control parameter, i.e., $\Omega_{i\tau}(k,\varphi)=\Omega_{i\tau}(k,\varphi+2\pi/q)$.
This mandates that the integral of the Berry curvature in Eq.~\eqref{eq:ChargeSpinPump} restricted to an adiabatic evolution $\varphi_0\to\varphi_0+2\pi/q$ is equal to a fraction $C_\tau/q$ of the Chern number, as shown in Ref.~\onlinecite{Marra2015}.
Therefore, the charge and the spin pumped during an adiabatic transformation $\Delta\varphi=2\pi m/q$ which is a multiple of the period $2\pi/q$ are given by
\begin{align}
Q|_{\,2\pi\frac{m}q}&=\frac{m}q(C_\up+C_\down),
\quad
\nonumber\\
S|_{\,2\pi\frac{m}q}&=\frac{m}q(C_\up-C_\down),
\label{eq:Fractions}
\end{align}
which gives a fractional charge and spin transferred if, e.g., $1<m<q$.
Hence, the charge and the spin pumped over a fraction of the pumping cycle is equal to a fraction of the charge and spin pumped over the whole adiabatic cycle.
Let us stress that such fractional quantization of the charge and spin transport is not due to the effect of interactions among particles, but it is solely due to the additional unitary symmetry of the Hamiltonian in the parameter space.
Hereafter we show an example of a system which exhibits a fractional quantization of the charge and spin transport.

\section{The model}

A possible route to achieve the fractional quantization of the charge and spin transport in a one-dimensional lattice is by the presence of a periodically amplitude-modulated and unidirectional Zeeman field which at each lattice site $n$ is given by
\begin{equation}
\mathbf{b}(n)=[b\cos(2\pi\alpha n+\varphi)+b_0]\hat{\mathbf{z}},
\label{eq:Zeeman}
\end{equation}
where $2\pi\alpha$, and $\varphi$ are respectively the wavevector and the phase-offset of the periodic modulation, while $b$ and $b_0$ are respectively the amplitude of the Zeeman field modulation and the intensity of a superimposed uniform field.
This system can be described by the Hamiltonian in momentum space $\mathcal{H}(\varphi)=\sum_k H(k,\varphi)$ with
\begin{align}
H(k,\varphi)=&\,
c_k^\dag
\!\cdot\!
(-2t\cos{k} + b_0\sigma_z)
\!\cdot\!
c_k^\nodag
+\nonumber\\&
+
\frac12{e^{\imath\varphi}}
c_k^\dag
\!\cdot\!
b\,\sigma_z
\!\cdot\!
c_{k+2\pi\alpha}^\nodag
+\text{h.c.},
\label{eq:Hamiltonian}
\end{align}
where the Nambu notation has been used with $c_k^\dag=[c^\dag_{k\up},c^\dag_{k\down}]$ and $c_k^\nodag=[c^\nodag_{k\up},c^\nodag_{k\down}]^\intercal$ the creation and annihilation spinors of states with momentum $k$ and $t$ the hopping parameter.
Hereafter we assume that the wavevector of the Zeeman field is commensurate to the lattice, i.e., that $\alpha=p/q$ is a rational number with $p,q\in\mathbb{Z}$ coprimes.
The presence of a finite Zeeman field explicitly breaks time-reversal symmetry in Hamiltonian~\eqref{eq:Hamiltonian}, which can be regarded as a spinfull generalization of the Harper-Hofstadter Hamiltonian\cite{Hofstadter1976,Harper1955}.
Notice that the periodicity of the Zeeman field is equivalent to a periodic potential acting separately on the two spin channels.
Indeed, for unidirectional Zeeman fields and in absence of spin-orbit coupling interactions, the two spin channels are decoupled and Hamiltonian~\eqref{eq:Hamiltonian} can be block-diagonalized into two independent Harper-Hofstadter Hamiltonians for each spin channel, i.e., $H(k,\varphi)=H_\up(k,\varphi)+H_\down(k,\varphi)$, where
\begin{align}
H_{\tau}(k,\varphi) =&\,
(-2t\cos{k} \pm b_0) c_{\tau k}^\dag
c_{\tau k}^\nodag
+\nonumber\\&
\pm
\frac12{e^{\imath\varphi}}
b \,
c_{\tau k}^\dag
c_{\tau k+2\pi\alpha}^\nodag
+\text{h.c.},
\label{eq:HamiltonianSCH}
\end{align}
where the $\pm$ sign corresponds respectively to the up and down spin channels.
Hamiltonian~\eqref{eq:HamiltonianSCH} coincides with the Harper-Hofstadter Hamiltonian\cite{Harper1955,Hofstadter1976} realized in a one-dimensional lattice with a periodically-modulated field with wavevector $2\pi\alpha$.
This Hamiltonian is formally equivalent to the Hamiltonian originally proposed by Hofstadter\cite{Hofstadter1976}, which describes a two-dimensional electron system in the presence of a strong magnetic field:
The wavevector $\alpha$ corresponds in this case to the magnetic flux per lattice cell in units of the flux quantum $h/e$.
Therefore the Chern numbers $C_\tau$ which label each of the $q-1$ intraband gaps of each spin channl separately, are given by the unique integer solution $|C_\tau|<q/2$ of the Diophantine equation\cite{Osadchy2001} $p\,C_\tau \equiv j_\tau \mod{q}$, where $j_\tau$ is the gap index with respect to the  Hamiltonian $H_\tau(k,\varphi)$.
Notice that the competition between the periodicity of the one-dimensional chain and of the Zeeman field gives rise to a superlattice with spatial periodicity greater than one unit cell.
As a consequence, the Brillouin zone and the energy levels are folded into a reduced Brillouin zone $[0,2\pi/q]$.

\begin{figure}
	\centering
\resizebox{1\columnwidth}{!}{%
	\includegraphics{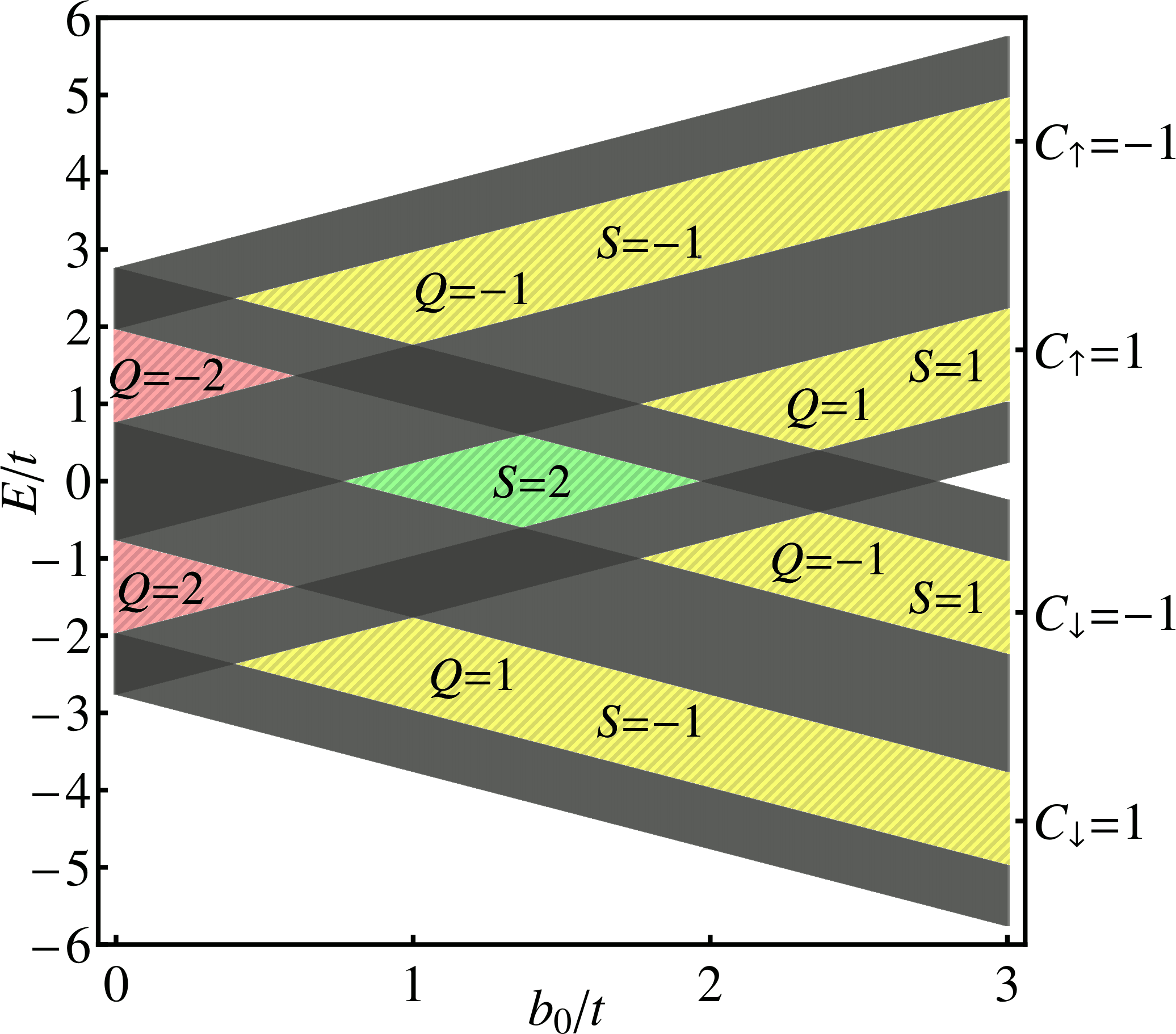}
}
	\caption{%
Energy levels (grey bands) of the two spin channels and intraband gaps (bright colors) as a function of the uniform Zeeman term $b_0$ and of the energy $E$ realized by a quantum pump with an amplitude-modulated Zeeman field with wavevector $2\pi\alpha=2\pi/3$ and intensity $b=2t$.
The intraband gaps are all topologically inequivalent and nontrivial, and are labeled by the charge $Q=C_\up+C_\down$ and the spin $S=C_\up-C_\down$ pumped during an adiabatic evolution of the phase-offset $\Delta\varphi=2\pi$.
For $2\pi\alpha=2\pi/3$ the system realizes a pure charge pump ($Q=\pm2$), a pure spin pump ($S=2$), or can pump charge and spin alternatively in the same ($Q=S=\pm1$) or in opposite directions ($Q=-S=\pm1$).
}
\label{fig:OverlappingGaps}
\end{figure}

\begin{figure}
	\centering
\resizebox{1\columnwidth}{!}{%
	\includegraphics{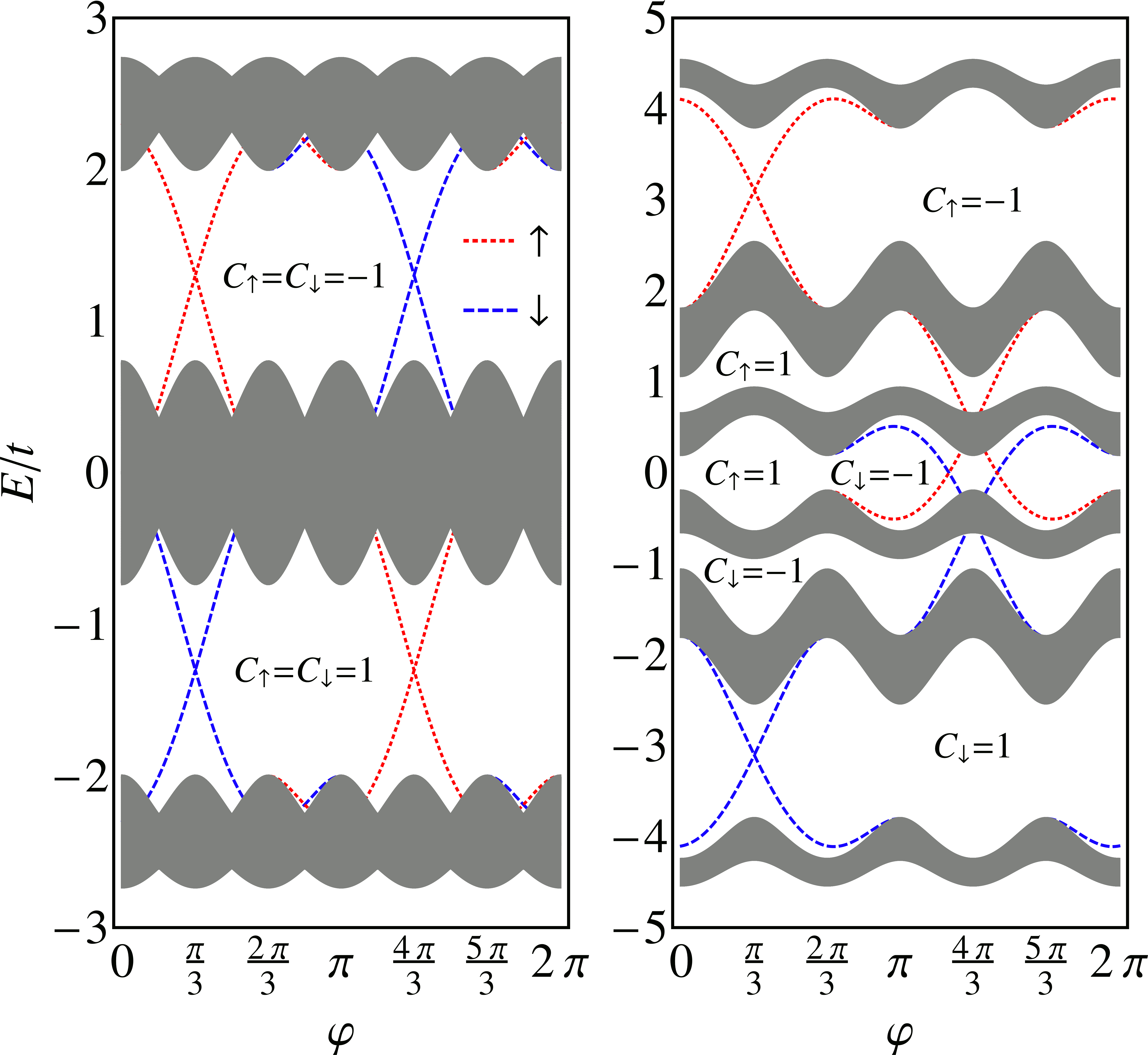}
}
	\caption{%
Bulk energy bands (grey) and edge states (bright colors) at the two boundaries of the system for each spin channel as a function of the phase-offset $\varphi$ of a quantum pump with an amplitude-modulated Zeeman field with wavevector $2\pi\alpha=2\pi/3$ and intensity $b=2t$, and with a vanishing $b_0=0$ (a) and with a finite uniform Zeeman term $b_0=1.8t$ (b) respectively.
The edge states have a well-defined spin, and for each spin channel $\tau=\up,\down$ the number of edge states at each boundary is equal to the corresponding Chern number $C_\tau$.
In the first case (a) each spin channel has the same number of edge states and same Chern numbers $C_\up=C_\down$, which corresponds to a pure charge pumping state with $Q=\pm2$ and $S=0$ realized in an infinite system (see Fig.~\ref{fig:OverlappingGaps}).
In the second case instead, the number of edge states and the Chern numbers differ $C_\up\neq C_\down$, which in an infinite system correspond to a pure spin pump with $S=2$ and $Q=0$ (central intraband gap around $E=0$) or in general to a net spin current $S\neq0$ (see Fig.~\ref{fig:OverlappingGaps}).
}
\label{fig:Edges}
\end{figure}

Remarkably, the uniform Zeeman term $b_0$ can be used to tune the topological properties of the system and, consequently, the amount of charge and spin pumped during an adiabatic cycle.
In fact, since the two spin channels are decoupled, the uniform Zeeman field $b_0$ is proportional to the energy splitting between energy levels with opposite spin and same band index, i.e., $E_{\up i}-E_{\down i}=2b_0$.
Consequently, the relative energy of the $q-1$ intraband gaps of each spin channel can be shifted in energy by tuning the uniform field $b_0$.
For instance, by increasing the field, the energies of the two spin channels move respectively upwards and downwards in energy.
Hence, the intraband gaps of the two spin channels can overlap in several different ways, giving rise to global gaps which can have different topological invariants $Q$ and $S$ (charge and spin Chern numbers).

\begin{figure*}
	\centering
\resizebox{1\textwidth}{!}{%
	\includegraphics{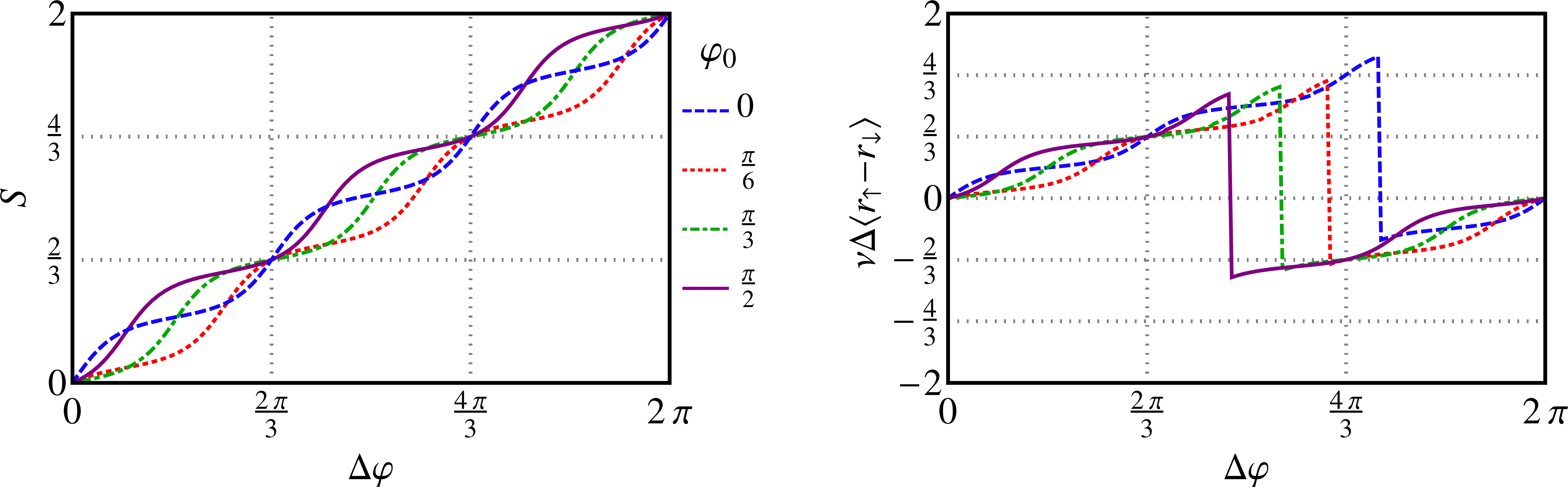}
}
	\caption{%
Amount of spin $S$ transferred (a) and variation of the spin dipole moment $\Delta\langle r_\up-r_\down\rangle$ (b) in a finite system of $L=165$ lattice sites, as a function of the variation of the phase-shift $\Delta\varphi$, of the amplitude-modulated Zeeman field with wavevector $2\pi\alpha=2\pi/3$ in the nontrivial state with $S=2$ and $Q=0$ ($b=2t$ and $b_0=1.8t$) at half filling.
Different lines correspond to different values of the initial phase-offset $\varphi_0$.
The spin pumped and the variations of the spin dipole moment at well-defined fractions of the pumping cycle $\Delta\varphi=2\pi m/q$ are fractionally quantized according to Eqs.~\eqref{eq:Fractions} and \eqref{eq:CenterOfChargeFractions}, independently from the initial value of the phase-offset $\varphi_0$.
}
\label{fig:Fractional}
\end{figure*}

Figure~\ref{fig:OverlappingGaps} shows the topological phase space of a quantum pump with amplitude-modulated Zeeman field with wavevector $2\pi\alpha=2\pi/3$ as a function of the uniform term $b_0$.
The energy bands (dark color) and the global gaps (bright colors) arise from the overlap respectively of the bands and the gaps of the two spin channels.
The global gaps are labeled by the charge $Q$ and the spin $S$ pumped during an adiabatic evolution of the phase-offset $\Delta\varphi=2\pi$.
The system exhibits a rich topological phase space, which includes inequivalent states realizing alternatively a pure charge pump (i.e., $Q\neq0$ and $S=0$), a spin pump (i.e., $Q=0$ and $S\neq0$), or a state which exhibits charge and spin pumping altogether (i.e., $Q\neq0$ and $S\neq0$).

As a consequence of the bulk-edge correspondence, nontrivial gaps are characterized by the presence of edge states at the boundary between the system and, e.g., a trivial insulating phase.
Since the spin channels are decoupled, these edge states are spin-polarized, i.e., they correspond to states with a well defined spin.
Figure~\ref{fig:Edges} shows the bulk energy bands and the edge states at the two boundaries of the system for each spin channel as a function of the phase-offset $\varphi$.
For each spin channel the number of edge states at each boundary is equal to the corresponding Chern number $C_\tau$.
The two panels correspond to different values of the uniform Zeeman term $b_0$.
The two intraband gaps in Fig.~\ref{fig:Edges}(a) with $C_\up=C_\down=\pm1$ correspond to the pure charge pumping state $Q=\pm2$ in Fig.~\ref{fig:OverlappingGaps}.
The central intraband gap in Fig.~\ref{fig:Edges}(b) with $C_\up=-C_\down=1$ correspond instead to the pure spin pumping state $S=2$, whereas the other four intraband gaps with $C_\down=\pm1$ and $C_\up=\pm1$ correspond to states with $Q\neq0$ and $S\neq0$ (both charge and spin are pumped) shown in Fig.~\ref{fig:OverlappingGaps}.

We notice that a similar system\cite{Mei2012,Zhang2016PLA} can be realized in a one-dimensional lattice with a spin-dependent potential given by
\begin{equation}
v_{\tau}(n)=v\cos(2\pi\alpha n\pm\varphi),
\label{eq:Counterpropagating}
\end{equation}
where the $\pm$ sign is for the up and down spin $\tau=\up,\down$ respectively.
Such a system can be also block-diagonalized into two copies of Harper-Hofstadter Hamiltonians respectively for each spin channel.
Differently from the case of a periodically modulated Zeeman field of Eq.~\eqref{eq:Zeeman}, this system does not break the time-reversal symmetry and consequently the charge Chern number is zero\cite{Mei2012}, being $C_\up=-C_\down$.
Therefore, the charge pumped over an adiabatic cycle vanishes $Q=0$, whereas the spin pumped is nonzero $S\neq0$ in the nontrivial phases where $C_\up=-C_\down=\pm1$.

\section{Fractional quantization of spin transport}

As discussed in Sec.~\ref{sec:general}, the charge and spin are fractionally quantized when Eq.~\eqref{eq:Unitary} is satisfied, i.e. if the Hamiltonian is periodic in the phase-offset $\varphi$ up to unitary transformations.
In particular, such unitary transformations correspond to a lattice translations.
It is easy to show from Eq.~\eqref{eq:HamiltonianSCH} that $H(k,\varphi+m 2\pi\alpha)=T(m) H(k,\varphi) T(-m)$ where $T(m)$ is the translation operator which translates the lattice by $m$ sites.
From this follows that\cite{Marra2015} the Hamiltonian $H(k,\varphi)$ is periodic in the phase-offset $\varphi$ with period $\Delta\varphi=2\pi/q$ up to a lattice translation, i.e.,
\begin{equation}
H(k,\varphi+m 2\pi/q)=T(C_m) H(k,\varphi) T(-C_m)
\end{equation}
where $T(C_m)$ translates the lattice by $C_m$ sites, with $C_m$ the integer solution of the Diophantine equation $p C_m\equiv m \mod q$.
In other words, changes of the phase-offset $\Delta\varphi=m2\pi/q$ (i.e., integer multiple of $2\pi/q$) are equivalent to a discrete lattice translation.
An analogous periodicity up to unitary transformations can be derived in the case of a one-dimensional discrete chain with a spin-dependent potential described by Eq.~\eqref{eq:Counterpropagating}.

Let us now verify numerically the fractional quantization of the spin transport, by calculating directly the spin pumped as the integral of the Berry curvature as a function of the variation of the phase-offset.
Figure~\ref{fig:Fractional}(a) shows the spin pumped during an adiabatic evolution of the system $\varphi_0\to\varphi_0+\Delta\varphi$ with $2\pi\alpha=2\pi/3$ in the nontrivial state with $S=2$ and $Q=0$, for different values of the initial phase-offset $\varphi_0$.
One can see that the amount of spin transferred is quantized as fractions of the total spin Chern number for $\Delta\varphi=2\pi m/q$ according to Eq.~\eqref{eq:Fractions}, independently from the initial value of the phase-offset $\varphi_0$.
The whole adiabatic cycle $\Delta\varphi=2\pi$ corresponds to a pumped spin equal to the total spin Chern number $S=2$.
We notice that the transferred spin is not quantized for $\Delta\varphi\neq2\pi m/q$.
The system exhibit also a fractional quantization of the charge pumped in the nontrivial phases with $Q\neq0$.

The fractional quantization of the charge pumped corresponds to an analogous fractional quantization of the variation $\Delta\langle r \rangle$ of the center of charge\cite{Marra2015}, which is defined in a finite system as $\langle r \rangle=\frac1{N}\int_0^L {\mathrm d}r \rho(r)r$, where $\rho(r)$ is the local density of states, $N$ the total number of particles, and $L$ the length of the system.
In a spinfull system, let us define the total center of charge and the `spin dipole moment' (i.e., the difference of the center of charge between the two channels)\footnote{The name `spin dipole moment' is used in analogy to the dipole moment of a charge distribution, which is given by \unexpanded{$\langle r_+ - r_- \rangle\equiv\langle r_+ \rangle - \langle r_- \rangle$} where \unexpanded{$\langle r_\pm\rangle$} are respectively the center of mass of the positive and negative charges of the distribution.}
respectively as
\begin{align}
\langle r_\up + r_\down \rangle&\equiv\langle r_\up \rangle + \langle r_\down \rangle
\nonumber\\
\langle r_\up - r_\down \rangle&\equiv\langle r_\up \rangle - \langle r_\down \rangle
\nonumber\\
\quad\text{where}\quad \langle r_\tau \rangle&=\frac1{N}\int_0^L {\mathrm d}r \rho_\tau(r)r
\label{eq:SpinDipoleMoment}
\end{align}
with $\rho_\tau(r)$ being the local density of states for each of the spin channels $\tau=\up,\down$.
As shown in Ref.~\onlinecite{Marra2015} for a spinless system, by integrating the continuity equation, one obtains that the variation of the total center of charge is equal to the charge pumped through the bulk up to an integer number.
By applying this argument separately to each spin channel, it follows that the variation of the total center of charge $\Delta\langle r_\up + r_\down \rangle$ and the variation of the `spin dipole moment' $\Delta\langle r_\up - r_\down \rangle$ over an adiabatic transformation $\Delta\varphi=2\pi m/q$ are given by
\begin{align}
\nu\Delta\langle r_\up + r_\down \rangle|_{\,2\pi\frac{m}q}&=\frac{m}q(C_\up+C_\down) \mod 1,
\nonumber\\
\nu\Delta\langle r_\up - r_\down \rangle|_{\,2\pi\frac{m}q}&=\frac{m}q(C_\up-C_\down) \mod 1,
\label{eq:CenterOfChargeFractions}
\end{align}
where $\nu=a N/L$ is the number of particles per lattice site ($a$ is the lattice parameter).
Notice that Eq.~\eqref{eq:CenterOfChargeFractions} is the analogous of Eq.~\eqref{eq:Fractions} for a finite system.

Figure~\ref{fig:Fractional}(b) shows the variations of the spin dipole moment $\Delta\langle r_\up - r_\down \rangle$ during an adiabatic evolution of the system $\varphi_0\to\varphi_0+\Delta\varphi$ with $2\pi\alpha=2\pi/3$ in the nontrivial state with $S=2$ and $Q=0$, for different values of the initial phase-offset $\varphi_0$.
The spin dipole moment is calculated directly from Eq.~\eqref{eq:SpinDipoleMoment} for an isolated and finite system ($L=165$ lattice sites) confined by a box-shaped potential.
The variations of the spin dipole moment are quantized as fractions of the total spin Chern number for $\Delta\varphi=2\pi m/q$ according to Eq.~\eqref{eq:CenterOfChargeFractions} up to an integer, and independently from the initial value of the phase-offset $\varphi_0$.

Notice that for a whole adiabatic cycle $\Delta\varphi=2\pi$, the variations of the center of charge and of the spin dipole moment always vanish, as a consequence of charge conservation in the system.
In other words, the system comes back to the original state ($\Delta\langle r_\up - r_\down \rangle=0$) at the end of the cycle.
For this reason, the shift of the center of charge due to the topological pumping has to be counterbalanced by an opposite change of the particle density at the edges of the system (cf.~Eq.~7 of Ref.~\onlinecite{Marra2015}).
This finite-size effect produces a discontinuous jump of the center of charge in systems with sharp boundary conditions (box-shaped potential), as in Fig.~\ref{fig:Fractional}(b).

\section{Experimental realization}

The experimental implementation of our proposal requires the realization of periodical and unidirectional Zeeman fields with a periodicity comparable with the underlying one-dimensional lattice.
Such kind of fields can be induced in nanowires by the presence of a contiguous antiferromagnetic material with a magnetic ordering vector which is commensurate with the lattice parameter of the wire, or via Moiré patterns realized by graphene on a ferromagnetic substrate\cite{Prezzi2014,Patera2015}.
Notice that periodically modulated Zeeman fields have been already considered in the context of topological quantum states in many proposals\cite{Klinovaja2012PRL,NadjPerge2013,Braunecker2013,Pientka2013,Klinovaja2013,Vazifeh2013,Kim2014,Arijit2014,Marra2016b}.
Moreover, this system can be implemented at a mesoscopic scale via artificial one-dimensional superlattices realized using nanolithographic design on ultrathin films\cite{Tadjine2016} or quantum dot solids\cite{Polini2013}, in the presence of nanomagnets.
However, the continuous control of the phase-offset $\varphi$ appears to be difficult to achieve in condensed matter.

On the other hand, topological spin pumps can be realized in optical lattices via ultracold fermionic atoms with two internal Zeeman states acting as pseudospins\cite{Celi2014,Cooper2015,Taddia2016}.
In these systems, a periodically modulated Zeeman field can be induced by a combination of radio frequency and optical-Raman coupling fields, which simultaneously couple the spin states of an ultracold Bose-Einstein condensate\cite{JimenezGarcia2012,Kartashov2013}.
Alternatively, the spin-dependent potential of Eq.~\eqref{eq:Counterpropagating} can be realized in optical lattices via two counterpropagating laser beams with linear polarization vectors\cite{Mei2012,Zhang2016PLA} forming an angle which corresponds to the phase-offset $\varphi$.
The fractional quantization of the spin transport can be verified by directly measuring the variation of the spin center of masses of the atomic cloud as a result of pumping by \emph{in situ} imaging\cite{Lohse2016,Nakajima2016}.

\section{Conclusions}

In this work we have described a novel property which can be observed in topological quantum pumps, i.e., the fractional quantization of the spin transport.
This phenomenon can be observed in the presence of a periodically modulated Zeeman field which is commensurate with the underlying lattice.
These systems can be realized in optical lattices of cold atoms, in nanowires in the presence of nanoengineered heterostructures, or in artificial one-dimensional superlattices.

\end{document}